\documentclass[%
 reprint,
 amsmath,amssymb,
 aps,
nofootinbib
]{revtex4-2}
\usepackage[utf8]{inputenc}

\usepackage{graphicx} 
\usepackage{dcolumn} 
\usepackage{bm} 
\usepackage{hyperref} 
\usepackage[export]{adjustbox}
\usepackage{xcolor}
\usepackage[normalem]{ulem}

\newcommand{\integral}{\int_{\mathcal{M}}\,d^4x\sqrt{-g}\,}
\newcommand{\average}[1]{\left\langle #1 \right\rangle}
\newcommand{\vac}{\mathrm{vac}}
\newcommand{\planck}[1]{M_P^{\,#1}}
\newcommand{\g}{g_{\mu\nu}}
\newcommand{\residual}{\Lambda_\mathrm{res}}

\begin{document}


\preprint{APS/123-QED}

\title{Exploring the self-tuning of the cosmological constant from Planck mass variation}

\author{Daniel Sobral Blanco}
 \email{Daniel.SobralBlanco@unige.ch}
\author{Lucas Lombriser}
 \email{Lucas.Lombriser@unige.ch}
\affiliation{Département de Physique Théorique, Université de Genève\\ 24 quai Ernest Ansermet, 1211 Genève 4, Switzerland}

\begin{abstract}
Recently, the variation of the Planck mass in the General Relativistic Einstein-Hilbert action was proposed as a self-tuning mechanism of the cosmological constant, preventing Standard Model vacuum energy from freely gravitating and enabling an estimation of the magnitude of its observed value.
We explore here new aspects of this proposal.
We first develop an equivalent Einstein-frame formalism to the current Jordan-frame formulation of the mechanism and use this to highlight similarities and differences of self-tuning to the sequestering mechanism.
We then show how with an extension of the local self-tuning action by a coupled Gauss-Bonnet term and a companion four-form field strength, graviton loops can be prevented from incapacitating the degravitation of the Standard Model vacuum energy.
For certain cases, we furthermore find that this extension can be recast as a Horndeski scalar-tensor theory and be embedded in the conventional local self-tuning formalism.
We then explore the possibility of a unification of inflation with self-tuning.
The resulting equations can alternatively be used to motivate a multiverse interpretation.
In this context, we revisit the coincidence problem and provide an estimation for  the probability of the emergence of intelligent life in  our Universe as a function of cosmic age, inferred from star and terrestrial planet formation processes.
We conclude that we live at a very typical epoch, where we should expect the energy densities of the cosmological constant and matter to be of comparable size.
For a dimensionless quantity to compare the emergence of life throughout the cosmic history of different universes in an anthropic analysis of the multiverse, we choose the order of magnitude difference of the evolving horizon size of a universe to the size of its proton as the basic building block of atoms, molecules, and eventually life.
For our Universe we find this number to form peak at approximately~42.
We leave the question of whether the same number is frequently assumed for the emergence of life across other universes or singles out a special case to future exploration.

\end{abstract}

\maketitle


\section{Introduction}

The fundamental nature underlying the cosmological constant $\Lambda$ in Einstein’s Theory of General Relativity continues to pose a tenacious enigma to modern physics. It is generally thought to be attributed to the gravitational effect of the vacuum energy that is anticipated to be of adequate magnitude to account for the observed late-time accelerated expansion of our Universe~\cite{Riess:1998cb,Perlmutter:1998np}. Quantum theoretical computations for this contribution, however, exceed measurements by $\gtrsim50$ orders of magnitude~\cite{Weinberg:1988cp,Martin:2012bt}. This may imply a missing prescription for the correct determination of the standard vacuum energy contribution. But it also motivates the conjecture of a yet undetermined mechanism~\cite{unruh,henneaux,henneaux2,barrow1,barrow2,kalpa1,kalpa5,lomb1,paper1} preventing vacuum energy from gravitating to full extent with cosmic acceleration attributed to a different origin such as a dark energy field permeating the Cosmos or a breakdown of General Relativity at large scales~\cite{Koyama:2015vza,Joyce:2016vqv,Ishak:2018his}. However, dark energy may need to be fine-tuned~\cite{Weinberg:1988cp} and the growing wealth of observations puts ever stronger constraints on alternatives to $\Lambda$~\cite{Aghanim:2018eyx,Troster:2020kai}, both on dark energy dynamics and the concept of cosmic self-acceleration from a modification of gravity that is confronted with hard challenges from the measured equality between the speeds of gravity and light~\cite{Monitor:2017mdv,Lombriser:2015sxa,Lombriser:2016yzn}. A curiosity of the cosmological constant, which perhaps may give a hint to finding a physical understanding, is that its energy density happens to be of comparable size to that of matter at the same time in cosmic history that we come into existence. This was not true in the past, nor will it be true in the future. This circumstance provokes a \emph{Why Now?} conundrum or the coincidence problem~\cite{Weinberg2,Martin:2012bt,Lombriser:2017cjy}. To address the different aspects of the \emph{Cosmological Constant Problem}, it can thus be useful to distinguish between the \emph{new} problem of the observed cosmic acceleration with a small observed coincident cosmological constant and the \emph{old} problem of vacuum gravitation that already arose before the discovery of the accelerated late-time expansion.

The severity and importance of the Cosmological Constant Problem has motivated a great amount of research (see Refs.~\cite{Padmanabhanb,nobbenhuisa,Polchinski,Bousso2,Padilla} for reviews). Symmetry principles have already been invoked early on as a potential solution to the old aspect, for instance, in the context of supersymmetry and scale invariance. Weinberg’s \emph{No-Go} Theorem~\cite{Weinberg:1988cp} offers here an important consistency check for any candidate solution to the old problem that invokes an extra dynamical scalar field in the matter sector. Typically, such a field could be introduced to ``absorb'' the large vacuum energy of the matter field sector, protecting the spacetime curvature accordingly. Then, assuming the vacuum state is symmetric under translations, such that the remaining symmetry is given by the group GL(4), Weinberg's theorem states that this is not possible without fine-tuning, except for the case of a universe with only massless particles, which is simply not the case for our Universe~\cite{Padilla}.
A relevant argument to solving both the old and new aspects of the Cosmological Constant Problem is furthermore given by Weinberg’s Anthropic Principle~\cite{Weinberg2} that only allows for values of $\Lambda$ that do not \emph{a priori} preclude the formation of structure and stars and hence the emergence of life and conscious life in particular.

There is a rich pool of proposed solutions to the Cosmological Constant Problem. Supergravity models, for example, typically involve the appearance of additional three-form gauge fields together with the Standard Model particles, which may turn the cosmological constant into a dynamical variable~\cite{aurilia,Brown1,Brown2,henneaux,henneaux2}. Perhaps the most successful candidate within this family of theories is the Bousso \& Polchinski neutralization mechanism~\cite{bousso}. It is a model with $J$ four-form gauge field strengths coupled to the correspondent $J$ charged membrane species that contribute to the effective value of the cosmological constant. The quantization of the gauge field fluxes yields a steady-step neutralization of the cosmological constant as long as the membrane species are continuously nucleating out of the vacuum. The mechanism may require $J\sim\mathcal{O}(100)$ membranes to operate, which agrees with the expected number arising from the compactification of String/M-theory to four dimensions~\cite{Bousso2,bousso}. Alternatively, one may, for instance, adopt a more usual path integral approach such as in the proposal of Barrow \& Shaw~\cite{barrow1,barrow2}. Using the partition function of the Universe, one can promote $\Lambda$ to be a variable parameter by summation over a given range of values. Taking a statistical approach one can then construct the probability distribution of observing a universe with a value $\Lambda' \in [\Lambda,\Lambda + d\Lambda]$. Anthropic selection also plays a role in this model in providing a window of allowed values of $\Lambda'$ with the observed value argued to be typical by the link of the time scale needed for galaxy formation and the age of the Universe. Other solutions propose the gravitational decoupling of the vacuum energy. An example of this is the sequestering mechanism~\cite{kalpa1,kalpa5}, in which in addition to the metric variation, $\Lambda$ and a matter coupling are varied in the Einstein-Hilbert action supplied with a global term that depends on the new variables. This introduces a global constraint equation that cancels the vacuum energy contributions at the level of the Einstein field equations. A local version can also be formulated, in which case the global term is replaced by two coupled four-form gauge field strengths.

Recently, another remedy of the Cosmological Constant Problem was proposed from the simple extra variation of the Einstein-Hilbert action with respect to the Planck mass or the gravitational coupling~\cite{lomb1}. An interpretation of this is the treatment of the Planck mass in the action as a global Lagrange multiplier that imposes General Relativistic dynamics on the metric prescribing the spacetime for the matter fields. The resulting additional constraint equation prevents vacuum energy from freely gravitating by the self-tuning of the classical $\Lambda$. Moreover, it was shown that an evaluation of the constraint equation for the observable Universe with a simple uniform prior on our location in its evolution yields an expected value for the current fractional energy density of the net radiatively stable cosmological constant of $\Omega_{\Lambda} = 0.704$~\cite{lomb1,paper1}, in good agreement with current measurements~\cite{Aghanim:2018eyx}. This simple approach therefore offers a solution to both the old and new aspects of the Cosmological Constant Problem. In Ref.~\cite{paper1} it was furthermore shown how similarly to sequestering the global self-tuning mechanism can be cast into a local form, specifically as a local scalar-tensor theory where the scalar field additionally couples to the flux of a three-form gauge field.

In this paper, we investigate further yet unexplored aspects of this new self-tuning mechanism. Specifically, we develop an alternative Einstein-frame formulation to the current Jordan-frame formalism of the mechanism. Furthermore, we address the question of whether graviton loops may spoil the degravitation of the Standard Model vacuum energy contributions. Finally, we explore a unification of inflation with the local self-tuning formalism and a related multiverse interpretation of the resulting equations. In this context, we also provide a crude estimate for the probability of the emergence of intelligent life in our Universe as a function of cosmic age, inferred from star and terrestrial planet formation processes.

The paper is organised as follows. Sec.~\ref{sec:selftuning} briefly reviews the previously established results on the self-tuning mechanism, including the global and local Jordan-frame formulations, the embedding of the local theory in Horndeski scalar-tensor gravity~\cite{Horndeski}, and the estimation of $\Omega_{\Lambda}$ from our likely location in cosmic history. In Sec.~\ref{sec:einsteinframe} we develop the Einstein-frame formulation of the self-tuning mechanism and we discuss in detail the similarities and differences between self-tuning and sequestering. In Sec.~\ref{sec:inflationary} we explore the unification of inflationary models with the self-tuning mechanism, conduct an expedition into the multiverse, and estimate the probability of the emergence of conscious life throughout cosmic history. Sec.~\ref{sec:gravitons} is devoted to a discussion of the effect of graviton loops on the self-tuning feature. Finally, we conclude our analysis in Sec.~\ref{sec:conclusions}.


\section{The self-tuning mechanism} \label{sec:selftuning}

Recently, the simple additional variation of the General Relativistic Einstein-Hilbert action with respect to the quadratic Planck mass $\planck{2}$ was proposed as a remedy to the Cosmological Constant Problem~\cite{lomb1}.
This allows the interpretation of the gravitational coupling as a Lagrange multiplier which imposes General Relativistic dynamics. 
The additional variation introduces a global constraint equation that yields a self-consistent self-tuning mechanism in which the large vacuum energy contributions to the cosmological constant are no longer freely gravitating.
A local counterpart to the global formalism was then developed in Ref.~\cite{paper1}.
We shall commence with a brief review of the local version in Sec.~\ref{sec:local} and then show how the global version emerges from that in Sec.~\ref{sec:global}.

\subsection{Local self-tuning in a simple scalar-tensor model} \label{sec:local}

The local self-tuning mechanism is best illustrated with the most basic scalar-tensor theory.
See, however, Sec.~\ref{sec:horndeski} for realisations in the most general scalar-tensor theories and Refs.~\cite{lomb1,paper1} for a discussion of how self-tuning may be realised in other theoretical frameworks.
Given a general manifold $\mathcal{M}$ equipped with a Lorentzian metric $g_{\mu\nu}$, the simplest set up for the self-tuning mechanism is the Jordan-frame scalar-tensor theory described by the action
\begin{equation}
\label{eq:1}
    S = \integral \left[\frac{\varphi}{2} R - V(\varphi) + \mathcal{L}_m(\g,\Psi_m)\right]
\end{equation}
with scalar field $\varphi$ and potential $V(\varphi)$.
Eq.~\eqref{eq:1} may represent the low-energy limit of a more fundamental theory, for instance, obtained from the four-dimensional compactification or reduction of a higher-dimensional theory of gravity~\cite{Jana:2020vov}.
$\mathcal{L}_m$ represents the matter content, e.g., the Standard Model of Particle Physics with $\Psi_m$ generically denoting any matter component.
We shall separate the vacuum energy contributions into the form $\mathcal{L}_m = \overline{\mathcal{L}}_m - \planck{2}\,\Lambda_\vac(\varphi)$, where $\Lambda_\vac(\varphi)$ may be regarded as an arbitrary function of $\varphi$ (see Ref.~\cite{paper1})\footnote{We could instead consider the term $M_P^2\,\Lambda_{vac}(\varphi)$ as part of the scalar potential.
Importantly, the resulting metric field equations and the conclusion that vacuum energy is not gravitating are the same regardless of this choice}.
Note that Eq.~\eqref{eq:1} is a Jordan-Brans-Dicke action with Brans-Dicke parameter $\omega = 0$, which is also the case for $f(R)$ theories. General Relativity is recovered in the limit $\varphi\,\rightarrow\,\planck{2}$.

In addition to the scalar-tensor action, we introduce the topological contribution
\begin{equation}
\label{eq:2}
    S_{\mathrm{A}} = \frac{1}{4!}\int_{\mathcal{M}}d^4x\, \epsilon^{\mu\nu\rho\sigma}\,\sigma(\varphi)\,F_{\mu\nu\rho\sigma} \,,
\end{equation}
where $F_{\mu\nu\rho\sigma} = \partial_{[\mu} A_{\nu\rho\sigma]}$ is the field strength of an auxiliary three-form gauge field $A_{\nu\rho\sigma}$ to which we couple the scalar field $\varphi$ through an arbitrary smooth, non-linear function $\sigma(\varphi)$.
Note that this additional sector may arise for example in supergravity models~\cite{aurilia,Brown1,Brown2,henneaux,henneaux2,bousso}. The local self-tuning mechanism is thus described by the total action $S + S_\mathrm{A}$.
The formulation is similar to the local sequestering framework of Ref.~\cite{kalpa5}
but with some important differences that we shall highlight in the following.
Most importantly, for self-tuning we only need one scalar and three-form gauge field rather than two of each, and we will see that here a classical counterterm is self-tuned to prevent vacuum energy from gravitating rather than obtaining a cancelation of the vacuum contributions.
Although we note that effectively the sequestering mechanism can be recovered in a special limit of the self-tuning framework.

The equations of motion for the total action $S+S_\mathrm{A}$ are obtained from its variations with respect to $g_{\mu\nu}$, $\varphi$, and $A_{\nu\rho\sigma}$. The variation with respect to the three-form gauge field yields the crucial condition $\partial_\mu\,\varphi = 0$ such that the dynamics of the scalar field is completely fixed, in the sense that it does not have any local propagating degrees of freedom.
Note that $\varphi$ shall be regarded as an auxiliary field. Hence, the presence of the additional $S_\mathrm{A}$ sector allows this auxiliary field to take the value $\planck{2}$ across the entire manifold $\mathcal{M}$. Alternatively, by integrating out the three-form sector from the beginning and taking the limit $\varphi\rightarrow\planck{2}$, we can interpret the gravitational coupling as a global Lagrange multiplier (Sec.~\ref{sec:global})~\cite{lomb1,paper1}.

Varying the total action with respect to $g_{\mu\nu}$ yields the Einstein field equations
\begin{equation}
\label{eq:3}
    \varphi\,G_{\mu\nu} + (V(\varphi) + \Lambda_\vac(\varphi))\,g_{\mu\nu} = \tau_{\mu\nu} \,,
\end{equation}
where $\tau_{\mu\nu}$ is the stress-energy tensor attributed to $\overline{\mathcal{L}}_m$.
Next, the variation of $S+S_\mathrm{A}$ with respect to $\varphi$ yields the constraint equation
\begin{equation}
\begin{split}
\label{eq:4}
    \integral\Bigg[\frac{1}{2} R - V'(\varphi) &- \planck{2} \Lambda_{\mathrm{vac}}'(\varphi) \\ &+ \frac{\sigma'(\varphi)}{4!}\frac{\varepsilon^{\mu\nu\rho\gamma}}{\sqrt{-g}} F_{\mu\nu\rho\gamma}\Bigg] = 0 \,,
\end{split}
\end{equation}
where primes denote derivatives with respect to $\varphi$ here and throughout the paper.
We stress that in Eqs.~\eqref{eq:3}--\eqref{eq:4} the terms $V(\varphi)$, $\Lambda_\vac(\varphi)$, $\sigma(\varphi)$, and their derivatives must be regarded as formal functions of the scalar field but that their value is fixed thanks to the condition $\partial_\mu\,\varphi = 0$.
This is why Eq.~\eqref{eq:4} is not a dynamical equation as would be the case for usual scalar-tensor theories but rather a constraint equation for each value of $\varphi$.

Taking the trace of Eq.~\eqref{eq:3} and splitting off an arbitrary classical counterterm $V_c(\varphi)$ from the potential with $\Delta\,V\equiv V-V_c$, Eq.~\eqref{eq:4} may be recast into the form
\begin{equation}
    \begin{split}
    \label{eq:5}
    (2-\beta) &\planck{-2} \Delta V \\
    &+ (2 -\alpha)(\Lambda_{\mathrm{vac}} + \planck{-2} V_c)  = \frac{\planck{-2}}{2}\langle\tau\rangle + \Delta\Lambda \,,
\end{split}
\end{equation}
where we have defined
\begin{equation}
 \label{eq:6}
    \frac{\planck{2}}{\varphi} \Delta\Lambda \equiv -\frac{\sigma'}{4!}\left\langle\frac{\varepsilon^{\mu\nu\rho\gamma}}{\sqrt{-g}} F_{\mu\nu\rho\gamma}\right\rangle = - \frac{\sigma'}{4!}\frac{\int d^4x \varepsilon^{\mu\nu\rho\gamma} F_{\mu\nu\rho\gamma}}{\int d^4x \sqrt{-g}} \,.
\end{equation}

For notational convenience, we have furthermore introduced here the set of parameters $\alpha\equiv \partial\,\ln{\Lambda_\vac}/\partial\,\ln{\varphi}$ and $\beta\equiv \partial\,\ln{\Delta\,V}/\partial\,\ln{\varphi}$ that characterize the dependency of $\Lambda_\vac$ and $\Delta\,V$ on the scalar field $\varphi$.
The counterterm can now be chosen such that the cancelation of $\Lambda_\vac$ becomes explicit. Hence, $\partial\,\ln{V_c}/\partial\,\ln{\varphi} \equiv \alpha$. Note that $\alpha$ and $\beta$ do not need to be constants but their values are fixed once we set the value of $\varphi$.
In Eqs.~\eqref{eq:5}--\eqref{eq:6} we have also introduced the braket notation $\average{...}=\integral(...)/\integral$ to denote the spacetime average of a quantity.
Note that such averages have been considered in the context of the Cosmological Constant Problem before such as in the Normalized General Relativity framework~\cite{davidson1,davidson2} (see also Refs.~\cite{tseytlin,arkanihamed,gabadadze}).

Finally, solving the constraint equation~\eqref{eq:5} for the classical counterterm $V_c$ and plugging the result into Eq.~\eqref{eq:3}, we obtain the effective Einstein equations for the self-tuning mechanism,
\begin{equation}
\begin{split}
    \label{eq:7}
 G_{\mu\nu} + \frac{1}{2-\alpha} \Bigg[(\beta&-\alpha)\planck{-2} \Delta V \\ &+ \frac{\planck{-2}}{2}\langle\tau\rangle + \Delta\Lambda\Bigg] g_{\mu\nu}  = \planck{-2} \tau_{\mu\nu} \,,
\end{split}
\end{equation}
where we have set $\varphi\equiv\planck{2}$.
Vacuum energy contributions are prevented from freely gravitating in these equations of motion, where $\Lambda \equiv \planck{-2}\,\Delta\,V$ is a free, radiatively stable, classical cosmological constant to be determined by measurement. Note that for $\alpha=\beta=1$ we recover the simplest global self-tuning result of Ref.~\cite{lomb1,paper1}, while the choice $\alpha=\beta=0$ reproduces the effective field equations of the sequestering mechanism of Refs.~\cite{kalpa1,kalpa5} but with a different expression for the $\Delta\Lambda$ term. In local sequestering, where one works with two additional three-form sectors, the quantity $\Delta\Lambda$ is essentially a ratio between the fluxes of both three-forms, whereas in Eq.~\eqref{eq:6} the denominator is instead the four-volume of the cosmic manifold $\mathcal{M}$. With the flux of the three-form gauge field in the numerator being a finite, small, ultraviolet-stable quantity, and assuming the Universe grows sufficiently old, it is natural to expect that $\Delta\Lambda\,\rightarrow\,0$.

\subsection{Self-tuning in the global limit} \label{sec:global}

The Einstein field equations~\eqref{eq:7} may also directly be obtained from varying the General Relativistic Einstein-Hilbert action both with respect to the metric and the quadratic Planck mass $\planck{2}$ and combining the results~\cite{lomb1}.
As we have seen in Sec.~\ref{sec:local} the global self-tuning formalism can be directly recovered from the local one after integrating out the three-form sector in the action and fixing $\varphi\equiv\planck{2}$ and $V(\varphi)\equiv \planck{2}\Lambda$ with $\Lambda$ being a free classical cosmological constant.
This reproduces the General Relativistic Einstein-Hilbert action with an additional sector $\sigma(\planck{2})\mathbf{F}$ sitting outside the integral, where $\mathbf{F} = (1/4!)\integral\,\varepsilon^{\mu\nu\rho\sigma}\,F_{\mu\nu\rho\sigma}$ is the flux of the three-form.
The separation of the Lagrangian density is performed as before, $\mathcal{L}_m = \overline{\mathcal{L}}_m - \planck{2}\Lambda_\vac$.
To illustrate the global self-tuning we adopt for simplicity a power-law scaling for the vacuum energy contributions such as $\Lambda_\vac = \planck{2\alpha}\,\overline{\Lambda}_\mathrm{vac}$, where the overbar denotes independency of $\planck{2}$, but we could instead also maintain a fully general behaviour in form of arbitrary functions of $\planck{2}$ (see Refs.~\cite{lomb1,paper1}).

The self-tuning is made explicit by splitting off a counterterm $\Lambda_c = \planck{2\alpha}\overline{\Lambda}_c$, analogously to the local case in Sec.~\ref{sec:local}, for which we then solve the constraint equation obtained from the variation of the total action with respect to $\planck{2}$.
The final expression for the effective Einstein equations is
\begin{equation}
    \label{eq:8}
    \begin{split}
        G_{\mu\nu} + \frac{1}{2-\alpha}\Bigg[(1-\alpha)\Lambda &- \overline{\mathbf{F}} \\
        &+\frac{\planck{-2}}{2}\average{\tau}\Bigg]g_{\mu\nu} = \planck{-2}\,\tau_{\mu\nu} \,,
    \end{split}
\end{equation}
where we have defined $\overline{\mathbf{F}} = (\sigma'\,\mathbf{F})/V_\mathcal{M}$, and $V_\mathcal{M}$ is the four-volume of the cosmic manifold.
Note that the additional constant contribution $\overline{\mathbf{F}}$ can simply be reabsorbed into the free classical $\Lambda$ by the redefinition $\Tilde{\Lambda} = \Lambda - \overline{\mathbf{F}}$. This contribution is radiatively stable and must be set by measurement.
Furthermore, the same line of reasoning as for $\Delta\Lambda$ holds for $\overline{\mathbf{F}}$ in that it is naturally expected to be a vanishing quantity.
Hence, we can simply drop the global sector in the global formalism. In other words, we can safely work with the General Relativistic Einstein-Hilbert action and simply perform the additional variation with respect to $\planck{2}$. The global self-tuning action and the General Relativistic Einstein-Hilbert action are equivalent.

\subsection{Self-tuning in Horndeski gravity} \label{sec:horndeski}

As mentioned in Sec.~\ref{sec:local} the local self-tuning mechanism is not limited to action~\eqref{eq:1} but can be realised for scalar-tensor theories more generally~\cite{paper1}. This can be shown by generalizing the procedure, for instance, to Horndeski gravity~\cite{Horndeski}, which describes the most general local scalar-tensor theory in four dimensions that yields at most second-order equations of motion. The Horndeski action can be written as~\cite{Horndeski,kobayashi}
\begin{equation}
    \label{eq:9}
    S = \integral \left[\frac{1}{2} \sum_{i=2}^5 \mathcal{L}_i(\g,\varphi)+\mathcal{L}_m(\g,\Psi_m)\right] \,,
\end{equation}
where the sum runs over the generalized Lagrangian densities
\begin{align}
    &\mathcal{L}_2 = G_2(\varphi,X) \,, \\
    &\mathcal{L}_3 = G_3(\varphi,X) \Box\phi \,, \\
    &\mathcal{L}_4 = G_4(\varphi,X) R + G_{4,X}(\varphi,X) \nonumber\\ 
    &\quad\qquad\qquad\qquad\quad\qquad \times\left[(\Box\varphi)^2+\varphi_{;\mu\nu} \varphi^{;\mu\nu}\right] \,, \\
    &\mathcal{L}_5 = G_5(\phi,X) G_{\mu\nu} \varphi^{;\mu\nu} \nonumber\\
    &- \frac{1}{6} G_{5,X}(\varphi,X)\left[(\Box\phi)^3 + 2\varphi_{;\mu}^\nu \varphi_{;\nu}^\alpha \varphi_{;\alpha}^\mu - 3\varphi_{;\mu\nu} \varphi^{;\mu\nu} \Box\varphi\right] \,.
\end{align}
Eq.~\eqref{eq:9} may again be seen as the effective low-energy limit of some more fundamental theory, for example, the four-dimensional compactification or reduction of a higher-dimensional theory of gravity~\cite{Jana:2020vov}.
The $G_i$ are general functions of the scalar field $\varphi$ with $X = -\,(1/2)\partial_{\mu}\varphi\, \partial^{\mu} \varphi$ denoting the canonical kinetic term. Semicolons indicate the standard covariant derivatives, $\varphi^{;\mu\nu} = \nabla^\mu\nabla^\nu\,\varphi$. Note that we recover the action~\eqref{eq:1} from \eqref{eq:9} for the choices $G_2 = -2\,V(\varphi)$, $G_4 = \varphi$, and $G_3=G_5=0$.

At inspection of the terms appearing in action~\eqref{eq:9}, we see that the local self-tuning mechanism remains operative once we add the additional sector with coupling of the scalar field $\varphi$ to the field strength of a three-form gauge field, Eq.~\eqref{eq:2}.
This term fixes the dynamics of $\varphi$ to take a constant value across the entire manifold $\mathcal{M}$. Therefore, this condition makes all the derivative terms in Eq.~\eqref{eq:9} vanish, leaving only a free choice of constants $G_2(\varphi)$ and $G_4(\varphi)$.
The local self-tuning mechanism can further be generalized to Degenerate Higher-Order Derivative Scalar-Tensor (DHOST) theories~\cite{langlois} or beyond-DHOST theories~\cite{Jana:2020vov}.
We refer the reader to Ref.~\cite{paper1} for a more detailed discussion on the operation of the self-tuning mechanism in more complex scalar-tensor models.

\subsection{The value of $\Lambda$} \label{sec:lambdaval}

Finally, a very interesting attribute of the self-tuning mechanism is that by the inherent relation of the cosmological constant to the global spacetime average of the matter content, it provides a natural framework to estimate the value of the observed cosmological constant $\Lambda$.
For illustration, let us consider the case $\alpha=\beta=1$ and work with a vanishing $\overline{\mathbf{F}}$ (or $\Delta\Lambda$).
The effective Einstein equations~\eqref{eq:8} or \eqref{eq:7} then contain a residual cosmological constant of the form $\Lambda_\mathrm{res} = \planck{-2}\average{\tau}/2$, which must correspond to the observed cosmological constant and is a calculable quantity~\cite{lomb2,lomb1,paper1}.

A na\"ive average over the cosmic background manifold for $\mathcal{M}$ in $\average{\tau}$ would suggest that $\Lambda_\mathrm{res}\rightarrow0$ for a universe that grows old.
In fact, that way the Universe should have collapsed one billion years ago for $\Lambda_\mathrm{res}$ to match the observed finite value of the cosmological constant~\cite{lomb1}.
However, due to cosmic acceleration, in the far future the observable Universe is not a simple extrapolation of our current cosmic background but instead reduces to the maximally gravitationally bound structure beyond which everything is redshifted away in finite time~\cite{krauss1,krauss2,adams}.
We can model the future Universe by adopting a halo model view of the Cosmos, in which one divides the cosmic manifold into the collection of ultimately isolated, maximally gravitationally bound, disconnected matter cells.
$\residual$ has been shown to then reproduce the observed value of the cosmological constant $\Lambda$ in each of the patches separately as well as in the empty space between them, no matter the sizes of these cells~\cite{lomb2,lomb1,paper1}.
Adopting the observable Universe as the manifold over which to perform the spacetime averages provides a self-consistent solution to the Cosmological Constant Problem.
However, since the properties of the manifold are determined by $\Lambda$, the value of $\Lambda$ remains a free quantity in this case.

Nevertheless, the relation of $\Lambda$ to $\average{\tau}$ in the self-tuning framework provides interesting insights into the \emph{Why now?} problem of why the energy densities of the matter and the cosmological constant happen to coincide today.
The evolution of $\tau$ in the observable Universe may be modeled by the cosmic background and the local spherical top-hat overdensity, which can be characterized in terms of its physical radius.
Its evolution is governed by the competition between the self-gravitation inside the overdensity and the background expansion.
The coincidence problem may then be phrased in terms of being located at a particular place in this evolution.
By then adopting a uniform prior on the dimensionless physical top-hat radius as the simplest ansatz to estimate our likely location (see Refs.~\cite{lomb2,paper1,lomb1}), we find a present fractional energy density in the cosmological constant of
\begin{equation}
\label{omega}
    \Omega_\Lambda(t_0)=0.704
\end{equation}
with $t_0$ denoting the current age of the Universe.
This value is in very good agreement with observations~\cite{Aghanim:2018eyx}.
We refer the reader to Refs.~\cite{lomb2,paper1,lomb1} for a discussion of more sophisticated priors in this anthropic ansatz.
In Sec.~\ref{sec:fortytwo} we shall perform a brief analysis of the emergence of intelligent life in our Universe.
A generalisation of this computation to arbitrary cosmological and fundamental parameters may serve to establish an anthropic theoretical prior in a multiverse approach in future work.

Finally, we note the interesting observation that the scenario discussed here with $\Lambda_\mathrm{res} = \planck{-2}\average{\tau}/2$ following from adopting $\alpha=\beta=1$ and $\overline{\mathbf{F}}=\Delta\Lambda=0$ reproduces the prediction for $\Lambda$ in the causal universe scenario of Refs.~\cite{gaztanaga1,gaztanaga2,gaztanaga3}.
An interesting observational implication of this scenario could be an inhomogeneous and anisotropic cosmological constant~\cite{Fosalba:2020gls}.
While the same is expected for the ultimate collapsed cells in the sequestering mechanism~\cite{lomb2}, the scenario discussed here for the self-tuning mechanism produces a homogeneous and isotropic $\Lambda$ (see, however, Sec.~\ref{sec:multiverse}).


\section{Einstein-frame formulation}\label{sec:einsteinframe}

So far the self-tuning mechanism has only been studied in the Jordan frame~\cite{lomb1,paper1}, where matter fields follow geodesics of the metric but the metric satisfies a modified Einstein equation.
We shall now examine how self-tuning operates in the conformally related and physically equivalent Einstein frame, where metrics satisfy the Einstein field equations but do not describe the geodesics of matter fields.
Hence, we consider the conformal transformation $\Tilde{g}_{\mu\nu} = \varphi\,g_{\mu\nu}$ of action~\eqref{eq:1}, which gives
\begin{equation}
\label{eq:14}
\begin{split}
        S = \int_\mathcal{M} d^4x\sqrt{-\Tilde{g}}\Bigg[\frac{\Tilde{R}}{2}&-\frac{3}{4}(\nabla_\mu\,\ln{\varphi})^2 - \varphi^{-2}V(\varphi) \\
        &\qquad\qquad+ \varphi^{-2} \mathcal{L}_m(\varphi\,\Tilde{g}^{\mu\nu},\Psi_m)\Bigg] \\
        +\frac{1}{4!}\int_\mathcal{M}d^4x\,&\varepsilon^{\mu\nu\rho\sigma}\sigma(\varphi)F_{\mu\nu\rho\sigma} \,,
\end{split}
\end{equation}
where tildes denote quantities in the Einstein frame.
We could furthermore redefine the scalar field $\varphi$ in order to recover a canonical kinetic term by using the relation $(d\Tilde{\varphi}/d\varphi)^2\equiv(3/2)(d\ln{\varphi}/d\varphi)^2$.

\subsection{Similarities with sequestering mechanism}

Before we discuss the self-tuning mechanism in the Einstein frame, we shall briefly discuss some similarities with the sequestering mechanism of Refs.~\cite{kalpa1,kalpa5}.
Here, the local version~\cite{kalpa5} is written in the Jordan frame whereas the global version~\cite{kalpa1} is in the Einstein frame.
The similarities between the sequestering and self-tuning mechanisms become more evident when we work in the global limit of Eq.~\eqref{eq:14}.
After integrating out the three-form sector, we can set the scalar field value to $\varphi\equiv\lambda^{-2}$, where $\lambda$ is a real parameter.
We can thus interpret $\lambda$ as the reciprocal of $\planck{}$ in the Einstein frame.
Taking into account the condition $\partial_\mu\,\varphi = 0$, the global limit of Eq.~\eqref{eq:14} reads
\begin{equation}
\label{eq:15}
\begin{split}
    S = \int_\mathcal{M}d^4x\sqrt{-\Tilde{g}}&\Bigg[\frac{\Tilde{R}}{2}-\lambda^4\,V(\lambda) \\
    &+ \lambda^{4}\,\mathcal{L}_m(\lambda^{-2}\,\Tilde{g}^{\mu\nu},\Psi_m)\Bigg] + \sigma(\lambda^{-2})\mathbf{F},
\end{split}
\end{equation}
where $\mathbf{F}$ is again the flux of the three-form on $\mathcal{M}$. This is clearly reminiscent of the global sequestering action of Ref.~\cite{kalpa1},
\begin{equation}
 \label{eq:sequestering}
\begin{split}
    S = \int_\mathcal{M}d^4x\sqrt{-\Tilde{g}}&\Bigg[\frac{\planck{2}}{2}\,\Tilde{R} - \Lambda \\
    &+ \lambda^4\,\mathcal{L}_m(\lambda^{-2}\,\Tilde{g}^{\mu\nu},\Psi_m)\Bigg] + \sigma\left(\frac{\Lambda}{\lambda^4\,\mu^4}\right).
\end{split}
\end{equation}
The sequestering action~\eqref{eq:sequestering} is varied separately with respect to $\Tilde{g}^{\mu\nu}$, $\Lambda$, and $\lambda$. 
In contrast, for self-tuning the role of the cosmological constant is played by the second term in Eq.~\eqref{eq:15}, which in this case depends on the parameter $\lambda$. 
However, this turns out to be the same parameter that controls the coupling of the matter sector.
Thus, if we redefine $\Tilde{\Lambda} \equiv \lambda^6\,V(\lambda)$ in Eq.~\eqref{eq:15} and rewrite the argument of the three-form coupling function, we could now perform variations with respect to $\lambda$ as well as with respect to $\Tilde{\Lambda}$ as is done for the global sequestering model.
However, as we shall see in the following the single extra variation with respect to $\lambda$ in the self-tuning recipe is sufficient for preventing vacuum fluctuations from freely gravitating.

\subsection{Self-tuning in the Einstein frame}

Performing the variation of the self-tuning Einstein-frame action~\eqref{eq:15} with respect to $\Tilde{g}^{\mu\nu}$, one finds the Einstein field equations
\begin{equation}
\label{eq:16}
    \Tilde{G}_{\mu\nu} + \lambda^4(V(\lambda) + \planck{2}\Lambda_\vac(\lambda))\,\Tilde{g}_{\mu\nu} = \Tilde{\tau}_{\mu\nu} \,,
\end{equation}
where $\Tilde{\tau}_{\mu\nu}$ is the stress energy tensor for the matter excitations, which is related to its counterpart in the Jordan frame through the relation $\Tilde{\tau}_{\mu\nu} = \lambda^2\,\tau_{\mu\nu}$.
This implies that the traces are related by the formula $\Tilde{\tau} = \lambda^4\,\tau$. Next, we shall perform the variation with respect to $\lambda$. This may be interpreted as the Einstein-frame analog of the $\planck{2}$ variation performed in Sec.\,\ref{sec:local}.
Analogously to our Jordan-frame calculations, we define the Einstein-frame parameters $\Tilde{\alpha}\equiv \partial \ln{\Lambda_\vac}/\partial\ln{\lambda} = \partial\ln{V_c}/\partial\ln{\lambda}$ and $\Tilde{\beta}\equiv \partial\ln{\Delta \tilde{V}}/\partial\ln{\lambda}$, where we have also introduced the arbitrary counterterm $V_c$, now a function of $\lambda$, which shares the same dependency as $\Lambda_\vac$. The constraint equation then reads
\begin{equation}
\label{eq:17}
    (4+\Tilde{\alpha})(V_c + \planck{2}\Lambda_\vac) + (4+\Tilde{\beta})\Delta \Tilde{V} = \lambda^{-4}\,\average{\tilde{\tau}} + \Tilde{\mathbf{F}} \,,
\end{equation}
where $\Delta \Tilde{V} \equiv V-V_c$ and $\Tilde{\mathbf{F}} = -(2/\lambda^6)\,\sigma_{\lambda}\,\mathbf{F}/\Tilde{V}_\mathcal{M}$ with $\sigma_{\lambda} \equiv d\sigma/d\lambda$ and $\Tilde{V}_\mathcal{M} = \lambda^{-4}\,V_\mathcal{M}$.
Again, although the different contributions are formal functions of $\lambda$, they take a constant value for any given value of the parameter.

Solving the constraint equation~\eqref{eq:17} for the counterterm $V_c$, and plugging the result into Eq.~\eqref{eq:16}, we immediately see that the vacuum energy contributions cancel out in the effective Einstein equations,
\begin{equation}
    \tilde{G}_{\mu\nu} + \frac{1}{4+\tilde{\alpha}}\left[(\tilde{\alpha}-\tilde{\beta})\lambda^4\Delta \tilde{V} + \average{\tilde{\tau}} + \lambda^4\,\tilde{\mathbf{F}}\right]\,\tilde{g}_{\mu\nu} = \tilde{\tau}_{\mu\nu} \,.
\end{equation}
Similar arguments to those given for the Jordan-frame result in Sec.~\ref{sec:selftuning} also follow here.
In particular, the effective sequestering mechanism is contained within this family of solutions for $\tilde{\alpha} = \tilde{\beta} = 0$, which gives the characteristic fraction of $1/4$ of $\average{\tilde{\tau}}$ as the contribution to the residual cosmological constant.
For $\tilde{\alpha} = -2$ we obtain a fraction of $1/2$. If furthermore $\tilde{\beta} = \tilde{\alpha}$, this corresponds to the choice of a potential of the form $V(\lambda) \sim \lambda^{-2}$. Note that we have fixed $\varphi \equiv \planck{2}$ in the Jordan frame whereas here we have used $\varphi \equiv \lambda^{-2}$.
The global parameters $\planck{}$ and $\lambda$ are thus interpreted as the reciprocal of each other. Hence, by construction, the dependency parameters in the different frames are also related such that $\tilde{\alpha} = -2\alpha$ and $\tilde{\beta} = -2\beta$.
The Einstein tensor is invariant, i.e., $\tilde{G}_{\mu\nu} = G_{\mu\nu}$ once we have fixed the scalar field to the aforementioned constant value.
The three-form flux contribution is simply $\tilde{\mathbf{F}} = -(2/\lambda^2)\,\mathbf{F}$.

Finally, with these considerations, we recover the Einstein field equations in Jordan frame, Eq.~\eqref{eq:7}, and physics is indeed independent of frame choice as expected.


\section{Inflation and the multiverse} \label{sec:inflationary}

It has been shown in Ref.~\cite{kalpa1} that an early inflationary era does not affect the residual cosmological constant and the vacuum energy cancelation in the sequestering mechanism since the spacetime averages are dominated by the late-time evolution.
The same conclusion applies to the self-tuning mechanism~\cite{lomb1,paper1}.
With a scalar field being the source of local self-tuning another immediate question that suggests itself is whether it could play the role of the inflaton.
This would imply a temporal division of the manifold $\mathcal{M}$ into an era where the field would be evolving and one where it is has become fixed.
Conceptually, vacuum energy may still be prevented from gravitating in the inflationary era due to a constraint equation imposed by the boundary to the post-inflationary era where the field is fixed, thereby avoiding Weinberg's \emph{No-Go} Theorem.
This scenario also opens another question, which is whether this boundary could also be drawn between causally disconnected regions of spacetime and how self-tuning works in a multiverse picture.
We shall briefly inspect these questions in Secs.~\ref{sec:inflaton}--\ref{sec:multiverse}.

\subsection{Can the scalar be the inflaton?} \label{sec:inflaton}

Eq.~\eqref{eq:1} is the action of a simple scalar-tensor theory, embedding for example $f(R)$ gravity, and hence it is perfectly suitable for describing an inflationary scenario. In this case, the scalar field $\varphi$ is the so-called inflaton,
for which one may impose particular conditions such as a slow-roll evolution.
Could we now conciliate both inflation and the self-tuning mechanism in one formalism with the inflaton field $\varphi$ driving inflation at early times and self-tuning through its late-time limit?

To address this question, let us modify the self-tuning action specified by Eqs.~\eqref{eq:1} and \eqref{eq:2} by including a step function in the topological sector,
\begin{equation}
\label{eq:20}
    \begin{split}
        S = \int_\mathcal{M}&d^4x\sqrt{-g}\left[\frac{\varphi}{2} R - V(\varphi) + \mathcal{L}_m(g^{\mu\nu},\Psi_m)\right] \\
            &+ \frac{1}{4!}\int_{\mathcal{M}}d^4x\, \Theta{\Big(\varphi-\planck{2}\Big)}\,\epsilon^{\mu\nu\rho\sigma}\,\sigma(\varphi)\,F_{\mu\nu\rho\sigma}.
    \end{split}
\end{equation}
Note that this is just a crude toy model. We suspect that a similar effect could for instance be achieved with a screening mechanism, which however lies beyond the scope of this work. By construction, the inflaton has active dynamics in the regime $\left(\varphi-\planck{2}\right)<0$, where the last term vanishes. In the complementary regime of $\left(\varphi-\planck{2}\right)\geq0$, we recover the self-tuning action of Eqs.~\eqref{eq:1} and \eqref{eq:2}.
This fixes the value of the inflaton at the scale $\planck{2}$ in the limit $\varphi\rightarrow\planck{2}$.
The impact of the modification in Eq.~\eqref{eq:20} on the equations of motion is minimal. In the limit $\varphi\rightarrow\planck{2}$, the topological contribution now takes the form $\Sigma(\varphi) \equiv \sigma(\varphi) + \sigma'(\varphi) \neq 0$.
This is forced to differ from zero for a scalar field with frozen dynamics, i.e., for the condition $\partial_\mu\,\varphi  = 0$ to hold. This is natural for a reasonable, non-linear differentiable function of the inflaton field, but notice that it removes the possibility that $\sigma(\varphi)$ takes the form of an exponential decay. Then, the constraint equation for the cancelation of vacuum energy contributions is completely analogous to that of Eq.~\eqref{eq:4} but with $\Sigma(\varphi)$ replacing $\sigma(\varphi)$.

From a formal point of view, we have consistently combined inflation with the self-tuning mechanism. The idea is essentially that the scalar field evolves following some inflationary model, possibly slow-rolling to the value $\varphi\rightarrow\planck{2}$. From this point, the inflaton dynamics freezes, the gravitational coupling is fixed and the self-tuning mechanism starts to operate, producing the degravitation of the vacuum energy contributions. This idea is, however, not free of difficulties. Further analysis is needed in order to determine if this scenario is phenomenologically feasible.
Furthermore, it would be interesting to analyse whether this could be further united within an extended model of the Higgs boson.

\subsection{Self-tuning in the inflationary era} \label{sec:inflation}

Given that with Eq.~\eqref{eq:20} the scalar field is dynamical at early times, we need to show that Weinberg's \emph{No-Go} Theorem is evaded and vacuum energy contributions are not freely gravitating during the inflationary era.
As we will see, this is achieved through a boundary constraint on the inflaton.
Formally, let us take Eq.~\eqref{eq:20} and split the spacetime manifold along the timelike direction in the region after the end of inflation $\mathcal{U}$ and the inflationary region $\mathcal{N} = \mathcal{M}\setminus\mathcal{U}$. By construction, we have $(\varphi-\planck{2})<0$ in $\mathcal{N}$ such that the scalar field $\varphi$ is dynamical and drives the expansion.
In general, each region will have different metric tensors, which shall be determined separately by solving the correspondent field equations. Let us denote the metric of $\mathcal{N}$ by $h_{\mu\nu}$ and that of $\mathcal{U}$ by $g_{\mu\nu}$. 

The region $\mathcal{U}$ is just the late-time description for the Universe specified by the action~\eqref{eq:1} with self-tuned cosmological constant. We thus recover the results from Sec.~\ref{sec:selftuning}. In the region $\mathcal{N}$, however, physics is described by a dynamical scalar-tensor theory, which for our choice of scalar-tensor action, Eq.~\eqref{eq:1}, gives the two equations of motion
\begin{equation}
\label{eq:21}
\begin{split}
    \varphi\,G_{\mu\nu}+ (V(\varphi)&+\planck{2}\,\Lambda_\vac(\varphi))\,h_{\mu\nu}\\
    & = (\nabla_\mu\nabla_\nu - h_{\mu\nu}\,\Box)\,\varphi + \tau_{\mu\nu} \,,
\end{split}
\end{equation}
\begin{equation}
\label{eq:22}
    \frac{R}{2} - V'(\varphi) - \planck{2}\,\Lambda_\vac'(\varphi) = 0 \,,
\end{equation}
where $G_{\mu\nu}$ and $R$ refer here to the metric $h_{\mu\nu}$.
Note that by virtue of Weinberg's \emph{No-Go} Theorem, the solution of these two equations alone will be a universe of massless particles or one in which we have not solved the fine-tuning problem for $\Lambda$~\cite{Weinberg:1988cp}.
We will see how this is evaded in the following.

We first recast Eq.~\eqref{eq:22} in the form of the usual Einstein field equations, isolating any differences to General Relativity into the term $\residual|_\mathcal{N}$,
\begin{equation}
\label{eq:24}
\begin{split}
    \Lambda_\mathrm{res}|_{\mathcal{N}}=\varphi^{-1}\Big[V(\varphi)+&\planck{2}\,\Lambda_\vac(\varphi)\\
    -&(h^{-1}_{\mu\nu}\nabla_\mu\nabla_\nu - \Box)\,\varphi\\
    -&(\tau_{\mu\nu}-\varphi\,\planck{-2}\,\tau_{\mu\nu})h^{-1}_{\mu\nu}\Big] \,,
\end{split}
\end{equation}
so that $G_{\mu\nu}+\residual|_\mathcal{N}\,h_{\mu\nu}=\planck{-2}\,\tau_{\mu\nu}$. Note that we have explicitly kept the $\varphi$ dependence of the arbitrary functions $V(\varphi)$ and $\Lambda_\vac(\varphi)$.
For simplicity, we shall assume power-law behaviours in $\varphi$ for $V_c$, $\Delta V \equiv V-V_c$, and $\Lambda_\vac$.
We could, however, adopt general functions of $\varphi$ analogously to Sec.~\ref{sec:selftuning}.
More specifically, we take $V_c = V_0\,\varphi^\alpha$, $\Lambda_\vac = \Lambda_0\,\varphi^\alpha$ and $\Delta V = \Delta V_0\,\varphi^\beta$,
where subscripts of zero denote amplitudes.
Note that these are smooth functions across the regions $\mathcal{N}$ and $\mathcal{U}$. For models where the initial conditions are such that $\varphi$ in $\mathcal{N}$ eventually reaches the value $\varphi=\planck{2}$, the field enters into the region $\mathcal{U}$ where its dynamics is frozen. Thus, $\residual|_\mathcal{\partial\mathcal{N}}$ must match the solution for $\residual|_\mathcal{U}$,
\begin{equation}
\label{eq:25}
    \Lambda_\mathrm{res}|_\mathcal{U} = \frac{\planck{-2}}{(2-\alpha)}\Bigg[(\beta-\alpha)\Delta V + \frac{1}{2}\average{\tau} + \planck{2}\,\Delta \Lambda\Bigg] \,.
\end{equation}
By taking the limit $\varphi\rightarrow\planck{2}$ in Eq.~\eqref{eq:24}, we have an expression for $\residual$ at the boundary of the region $\partial\mathcal{N}$,
\begin{equation}
    \begin{split}
        \residual|_{\partial\mathcal{N}}=\planck{-2}\Big[\Delta V_0\,\planck{2\beta} &+ V_0\,\planck{2\alpha} + \Lambda_0\,\planck{2\alpha+2}\\
        &-(h^{-1}_{\mu\nu}\nabla_\mu\nabla_\nu - \Box)\varphi|_{\partial N}\Big] \,.
    \end{split}
\end{equation}
Using the constraint~\eqref{eq:5}, this becomes
\begin{equation}
\begin{split}
        \residual|_{\partial\mathcal{N}}=\frac{\planck{-2}}{2-\alpha}\Bigg[ (\beta & -\alpha) \Delta V +\frac{1}{2}\average{\tau}_{\mathcal{U}}+\planck{2}\,\Delta\Lambda\\
        &-\planck{-2}(h^{-1}_{\mu\nu}\nabla_\mu\nabla_\nu - \Box)\varphi|_{\partial N}\Bigg].
\end{split}
\end{equation}
By comparison with Eq.~\eqref{eq:25}, there is an extra contribution from the dynamics of the scalar field at $\mathcal{\partial N}$.
For a smooth transition of $\varphi$ to $M_P^2$, perhaps as a form of attractor, slow roll, or the condition on the second derivatives introduced in Sec.~\ref{sec:horndeski}, hence, $\residual|_\mathcal{\partial N}=\residual|_\mathcal{U}$.

\subsection{Expedition into the multiverse} \label{sec:multiverse}

The modified self-tuning action in Eq.~\eqref{eq:20} opens the door to a multiverse interpretation.
More specifically, we may consider a Cosmos that is divided into several spacetime patches, each an own universe.
This essentially corresponds to the late-time scenario predicted by Eternal Inflation~\cite{Guth}.
Some of them may see a dynamical scalar field $\varphi$ whereas others may have a frozen scalar field.
This is similar to the inflationary scenario discussed in Sec.~\ref{sec:inflation}, however, with the manifold now not split timelike but spacelike.
Note that we already perform a spacetime division of the background Universe by adopting the observable Universe for $\mathcal{M}$ in Sec.~\ref{sec:selftuning}.
Self-tuning would operate in universes with fixed scalar field dynamics, preventing vacuum energy contributions from freely gravitating and proportioning a residual cosmological constant self-consistent with the magnitude of the accelerated expansion observed in this patch.
Generally, we could consider at least two types, or levels, of multiverses~\cite{Tegmark:2003sr}:
\begin{itemize}
    \item \textbf{Type-I:} Different regions will have different initial conditions with varying configurations of the matter content. The dynamics of the scalar field is therefore different across the different regions but the fundamental constants are fixed. 
    \item \textbf{Type-II:} Fundamental constants such as the Planck mass may also vary across the different regions. 
\end{itemize}

We shall briefly inspect a simple multiverse picture for self-tuning.
For this, consider a simple set up where we have only two disconnected regions.
Let us denote by $\mathcal{U}$ the region of the spacetime manifold $\mathcal{M}$ where the condition $(\varphi-\planck{2})\geq 0$ holds in Eq.~\eqref{eq:20} such that the self-tuning mechanism operates in $\mathcal{U}$.
In the complementary region, denoted by $\mathcal{N} = \mathcal{M}\setminus\mathcal{U}$, we shall have $(\varphi-\planck{2})<0$ such that the scalar field is dynamical.
Note that the region $\mathcal{N}$ may also well represent the collection of a set of many other spacetime patches different from $\mathcal{U}$. 
Action~\eqref{eq:20} can now be split into two parts, which in general will have different metrics. Let us call $h_{\mu\nu}$ the metric in the region $\mathcal{N}$ and $g_{\mu\nu}$ the metric in the self-tuning region $\mathcal{U}$.
They are determined by the correspondent Einstein equations in each region.
For simplicity, we shall furthermore assume that the matter content is the same in both patches and that $\Lambda_\vac|_\mathcal{N} = \Lambda_\vac|_\mathcal{U}$.
The results of Sec.~\ref{sec:inflation} are formally valid for this new set up but they are given a new interpretation here.
In particular, by solving the dynamics with the constraint $\residual|_{\partial\mathcal{N}}=\residual|_\mathcal{U}$, we can find a model that allows for regions of dynamical $\varphi$ while also canceling vacuum energy contributions in those regions.

\subsubsection*{Variation of Fundamental Constants}

Let us briefly inspect how a Type-II (or Level II) multiverse could arise in the framework of self tuning.
We can again start with a Horndeski scalar-tensor action as in Eq.~\eqref{eq:9} or beyond~\cite{langlois,Jana:2020vov} but instead consider the matter sector
\begin{eqnarray}
 S_m & = & \sum_i \int d^4x \sqrt{-g} B_i(\phi_i,\chi_i) \times \nonumber\\
 & & \mathcal{L}_{m_i}\left(\Psi_{m_i};A_i^2(\Phi_i,\xi_i)g_{\mu\nu} + D_i^2(\Phi_i,\xi_i)\partial_{\mu}\Phi_i\partial_{\nu}\Phi_i\right) \nonumber\\
 \label{eq:fundconst1}
\end{eqnarray}
and the three-form sector
\begin{eqnarray}
 S_A & = & \frac{1}{4!} \int d^4x \epsilon^{\mu\nu\rho\sigma} \left( \sigma(\varphi) F_{\mu\nu\rho\sigma} + \sum_i \sigma_i(\phi_i) (F_i)_{\mu\nu\rho\sigma} \right. \nonumber\\
 & & \left. + \hat{\sigma}_i(\Phi_i) (\hat{F}_i)_{\mu\nu\rho\sigma} \right) \label{eq:fundconst2}
\end{eqnarray}
with an additional sector $S_{\phi_i,\Phi_i}$ for the additional scalar fields $\phi_i$, $\Phi_i$, where $\chi_i \equiv -\partial_{\mu}\phi_i\partial^{\mu}\phi_i/2$ and $\xi_i \equiv -\partial_{\mu}\Phi_i\partial^{\mu}\Phi_i/2$.
We could also have cross-terms such as a potential $V(\varphi,\phi_i,\Phi_i)$.

Action~\eqref{eq:fundconst1} breaks the weak equivalence principle in case of multiple fields and non-universality of the couplings to different matter species.
An evolution of $B_i$ implies an evolution of the coupling of $\phi_i$ to the matter species $i$.
For instance, for the electromagnetic Lagrangian $\mathcal{L}_{m_i} = \mathcal{L}_{\rm em}$ this leads to an evolving fine-structure constant $\alpha$ (e.g., Ref.~\cite{Hees:2014lfa}).
$A_i$ and $B_i$ introduce a species-dependent conformal and disformal coupling to the metric $g_{\mu\nu}$.
This can lead to an evolution of particle masses and propagation speeds of the respective interactions (e.g.,~\cite{Domenech:2016yxd,vandeBruck:2015rma}).
Examples for such couplings include interactions between dark energy and dark matter or self interactions.
$A_i=A_j=B_i^{1/4}$, $D_i=0$ $\forall i,j$ corresponds to a scalar-tensor theory in Einstein frame and $B=A^4=M_P^{-2}G_4(\varphi)$, $D_i=G_5=G_5=0$ with $\phi_i=\varphi$ recovers General Relativity.

Adding the three-form sector~\eqref{eq:fundconst2} renders these couplings, masses and speeds constant as they fix the respective scalar fields on the domain.
As in Eq.~\eqref{eq:20}, one can introduce domains of evolving fundamental parameters by performing the analoguous replacement
\begin{eqnarray}
    & & \frac{1}{4!} \int d^4x \epsilon^{\mu\nu\rho\sigma} \sigma(\varphi) F_{\mu\nu\rho\sigma} 
    \nonumber\\
    & & \quad\quad \rightarrow \frac{1}{4!} \int d^4x \Theta(\varphi-\varphi_0) \epsilon^{\mu\nu\rho\sigma} \sigma(\varphi) F_{\mu\nu\rho\sigma}
\end{eqnarray}
for the scalar fields $\phi_i$ and $\Phi_i$.
The evolving fundamental parameters may also freeze to different constant values between different domains.
In frozen domains, one can then also formulate a global version of this picture with the action
\begin{eqnarray}
 S & = & \frac{M_{P}^2}{2} \int d^4x \sqrt{-g}\left( R
 - 2 \Lambda(M_{pl}^2,\alpha_i,m_i)\right)
 \nonumber\\
 & & + \sum_i \int d^4x \sqrt{-g} B_i(\alpha_i) \mathcal{L}_{m_i}\left(\psi_{m_i};A_i^2(m_i)g_{\mu\nu} \right) \,,
\end{eqnarray}
where one could furthermore allow for variation in the speeds $c_i$ with the introduction of disformal couplings $D_i$.

\subsection{Life, the Universe and Everything} \label{sec:fortytwo}

A multiverse picture coupled with the anthropic principle offers an interesting approach to understanding the material configuration of our Cosmos or even the fundamental constants and laws of Nature.
To understand the emergence of life, we must understand fundamental physics, but to understand fundamental physics, we must understand the emergence of life.
As discussed in Sec.~\ref{sec:lambdaval}, the self-tuning mechanism shows a natural display of anthropic aspects, where we have used a prior on our location in the cosmic history to estimate the observed current fractional energy in the cosmological constant $\Omega_{\Lambda}$.
In the following, we shall perform a more involved yet still rudimentary estimate for the formation probability of conscious observers in the cosmic history of our Universe.
We will infer this from the star formation rate, the evolution of cosmic metallicty, and, depending on these, the formation probability of terrestrial planets as potential hosts of intelligent life.

We first need to compute the star formation history.
For this we follow Ref.~\cite{Madau:2014bja}, where we also adopt the same cosmological parameters.
Specifically, we use $\Omega_m=0.3$, $\Omega_b=0.045$, and $h=0.7$ for the fractional energy densities in total matter and baryons and the dimensionless Hubble parameter, respectively.
Furthermore, we adopt the solar metallicity $Z_{\odot}=0.02$, where $Z$ more generally denotes the metallicity in the gas and newly formed stars.
We use $R=0.27$ for the return fraction, quantifying the mass fraction of each generation of stars that is returned to the interstellar and intergalactic medium.
Finally, we adopt $y=0.019$ for the net metal yield, the fractional mass of new heavy elements created and deposited in the interstellar and intergalactic medium by each generation of stars.
%
\begin{figure*}[ht!]
\centering
 \resizebox{0.477\textwidth}{!}{
 \includegraphics[valign=t]{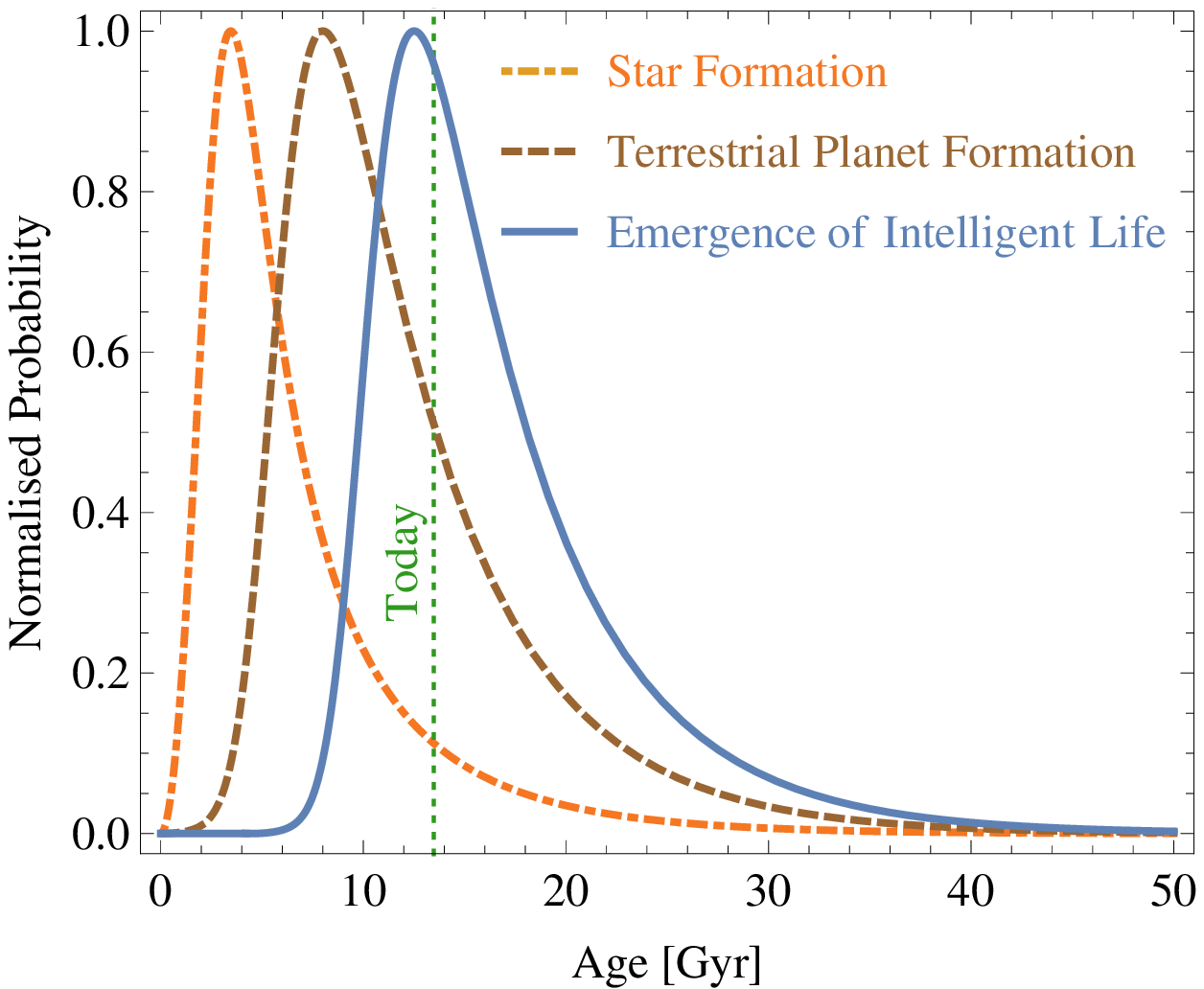}
 }
 \resizebox{0.477\textwidth}{!}{
 \includegraphics[valign=t]{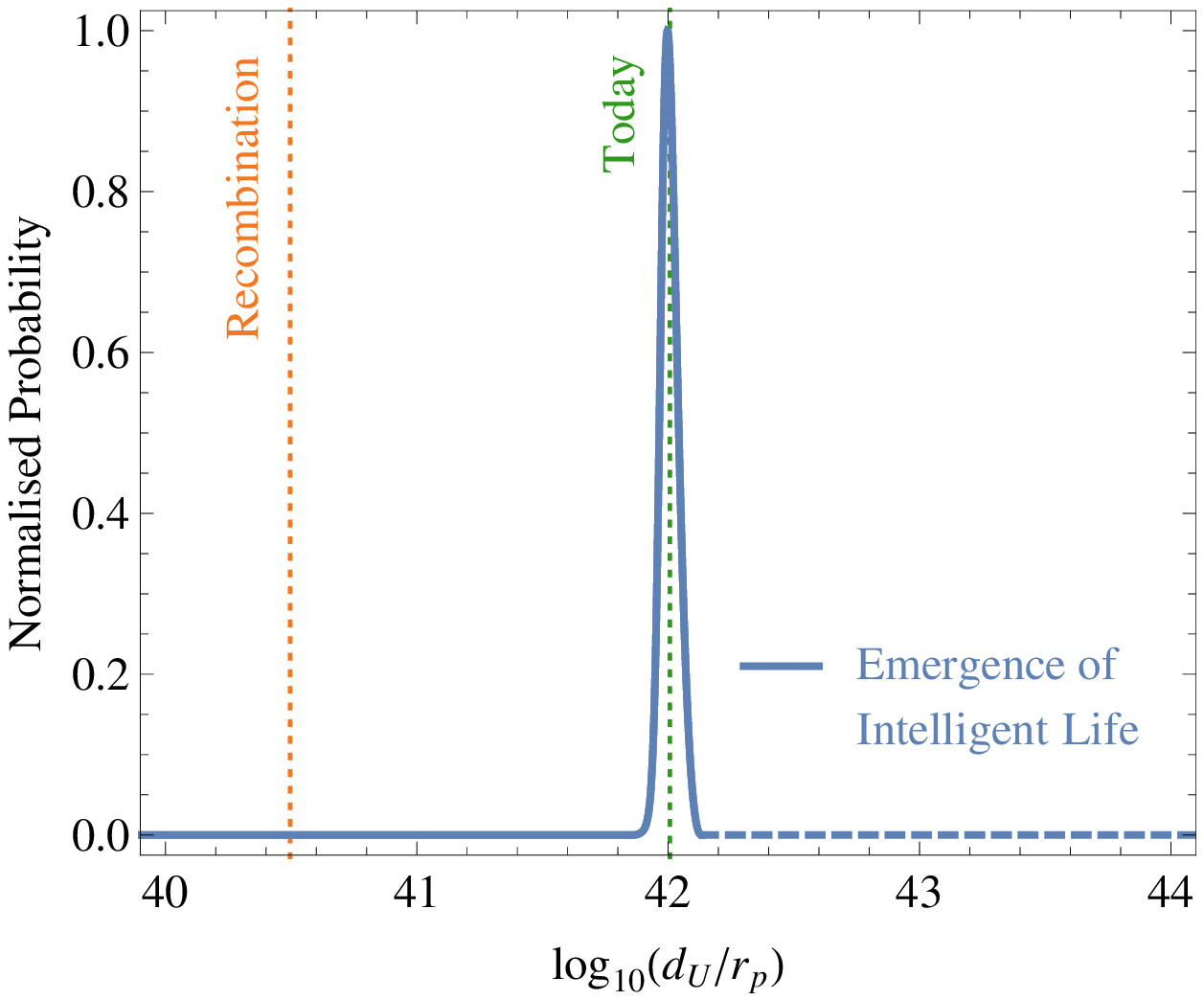}
 }
\caption{Normalised probabilities for star formation, formation of terrestrial planets, and emergence of intelligent life. \emph{Left}:~Probabilities as a function of the age of our Universe. \emph{Right}:~Probability of the emergence of life as a function of a dimensionless measure of cosmic time, here the orders of magnitude difference between the evolving size of the Universe to the particle horizon normalised to the size of the proton as the basic building block of atoms, molecules, and eventually life.
}
\label{fig:emergenceoflife}
\end{figure*}
%

The star formation rate is empirically given by~\cite{Madau:2014bja}
\begin{equation}
 \frac{d\rho_*}{dt} = \frac{0.015(1-R)(1+z)^{2.7}}{1+[(1+z)/2.9]^{5.6}}~{\rm M_{\odot} yr^{-1} Mpc^{-3}} \,,
\end{equation}
where $\rho_*$ is the cosmological energy density in stars, including remnants.
The cosmic metallicity is determined by $Z_b(z) = y \rho_*(z)/\rho_b$, where $\rho_b$ is the total baryonic matter density.
The formation of terrestrial planets can be modelled following Ref.~\cite{Zackrisson:2016wdz} with a formation probability of $P_{\rm FTP} \propto k(Z)$, where $k(Z)=(Z-10^{-4})/(10^{-3}-10^{-4})$ interpolates between 0 and 1.
For simplicity we shall adopt $Z \approx Z_b$.
Finally, for the time span between the formation of a terrestrial planet and the emergence of conscious life forms as observers, we shall adopt 4.5 billion years, motivated by our own evolution on Earth.
Combining these relations, we can determine the observer formation probability, or the emergence probability of intelligent life, throughout cosmic history.
We show our results in Fig.~\ref{fig:emergenceoflife}.
An important conclusion that can be drawn from these results is that from an anthropic standpoint there is no coincidence or \emph{Why Now?} problem as we live close to the observer formation peak.

Finally, we may ask ourselves the question whether life may form in other patches of the multiverse with different configurations of cosmological or even fundamental parameters.
In that case, rather than working with timescales that differ between different universes, we should opt for a sensible dimensionless parameter to describe the evolution of a universe and its relation to the formation of life.
There are many options available.
One choice one can take is to quantify this evolution in terms of a comparison of the evolving size of the cosmological particle horizon to the size of the proton as the basic building block of atoms, molecules, and eventually life.
More specifically, this compares the maximal Compton wavelength of a particle that could still be resolved within the horizon to the proton size.
We show the emergence probability for life as a function of this quantity, corresponding to the ratio of the diameter to the particle horizon $d_U$ to the proton radius $r_p$ in Fig.~\ref{fig:emergenceoflife}.
The observer formation probability peaks close to 42 orders of magnitude difference between the two.
An interesting question would now be to determine this number (or that of other dimensionless measures of cosmic time) for different universes with different cosmological and fundamental parameters.
We leave it to future work to investigate whether the emergence of life in the multiverse always peaks around 42
or if this singles out a special case, and hence determine whether this number in some sense may indeed qualify well as the answer to the Ultimate Question of Life, the Universe and Everything~\cite{HG2G}.


\section{Graviton loops}\label{sec:gravitons}

Finally, we shall conclude our exploration of novel aspects of the self-tuning mechanism in this paper with an analysis of the effect of graviton loops on the self-tuning property.
It has been shown in Refs.~\cite{lomb1,paper1} that the self-tuning mechanism is a consistent semi-classical formalism, very similar to what is the case for the sequestering scenario (see Refs.~\cite{kalpa1,kalpa5}).
In the semi-classical regime, one considers the gravitational field as classical, while the matter sector is treated quantum-mechanically. Then, quantum corrections give contributions to the cosmological constant. However, if we also consider quantum effects in gravity, we will also find corrections to the gravitational coupling. In particular, 1PI matter and graviton loops modify the cosmological constant, while pure gravity loops are responsible for the corrections to the gravitational coupling. An exhaustive analysis of the different effects of the quantum corrections and the ultraviolet sensitivity of sequestering, both in the Jordan and Einstein frames, is presented in Ref.~\cite{seerylawrence}.
In the case of the sequestering model, it has been shown that to cope with graviton loops the local model must be extended~\cite{kalpa6}.
We shall briefly review this discussion and explore its extension to the self-tuning mechanism.

\subsection{The Gauss-Bonnet invariant} \label{subsec:invariant}

The fact that loops involving graviton virtual lines are explicitly Planck mass dependent contributions invalidates the vacuum energy sequestering in its original formulation~\cite{kalpa5,kalpa6}. Recall that we are required to perform a variation with respect to the gravitational coupling, i.e., $\planck{2}$ (or $\kappa^2$ in Ref.~\cite{kalpa6}).
Similarly, the self-tuning in its original form is expected to fail in the presence of graviton loops.
In sequestering, the variation with respect to $\planck{2}$ in the global limit is such that $\average{R}$ is fixed by a radiatively stable ratio of the two three-form fluxes introduced in the formalism, which is just a linear relationship between $\average{R}$ and a combination of the three-form fluxes. For self-tuning, this instead involves a single three-form flux.
As was shown in Ref.~\cite{kalpa6} any other curvature invariant that is not constructed purely out of the Weyl tensor and the traceless part of the Ricci tensor turns out to take care of the sequestering of the graviton loops.
This yields a constraint that is a polynomial in the difference $\Lambda-\average{T}/4$. Moreover, when the variational constraint is not explicitly dependent on $\planck{2}$, all of the Planck mass dependent contributions cancel from the residual energy-momentum tensor, at least to leading order \cite{kalpa6,seerylawrence}.

The minimal candidate to accomplish this task is the Gauss-Bonnet curvature invariant, defined by $R_\mathrm{GB} = R^2_{\mu\nu\rho\sigma} - 2\,R^2_{\mu\nu} + R^2$. The contribution of this term to the action is trivial in four dimensions as it can be recast as a total derivative that we can integrate to zero.
In this sense, the addition of $R_\mathrm{GB}$ to the action merely introduces a topological modification that does not affect the local physics at finite wavelength. The fact that (the square of) the Ricci scalar appears explicitly in its definition is what ensures that a constraint is obtained that picks the correct counterterms to sequester all the next-order curvature corrections, i.e., the leading-order graviton loops.
Hence, the general local sequestering action reads~\cite{kalpa6}
\begin{equation}
\label{eq:33}
\begin{split}
        S = \integral \Bigg[\frac{\kappa^2(x)}{2} R &+ \theta(x)\,R_\mathrm{GB} \\
        &- \Lambda(x) + \mathcal{L}_m(g^{\mu\nu}, \Psi_m)\Bigg] \,,
\end{split}
\end{equation}
which includes the $R_\mathrm{GB}$ term coupled to a local field $\theta(x)$.
The topological sector $S_A$ of this theory consists of the term $\sigma(\Lambda/\mu^4)\,F_{\mu\nu\rho\sigma}$ as well as two more coupled three-form gauge-fields, one coupled via a function of $\theta(x)$, i.e., $\hat{\sigma}(\theta)\,\hat{F}_{\mu\nu\rho\sigma}$, and the other via a function of $\kappa^2(x)$, i.e., $\tilde{\sigma}(\kappa^2(x)/\planck{2})\tilde{F}_{\mu\nu\rho\sigma}$.
The functions $\kappa^2(x)$, $\Lambda(x)$, and $\theta(x)$ are again auxiliary rather than ordinary fields.
The variations are now performed with respect to $g_{\mu\nu}$, the three gauge fields as well as the three auxiliary fields. Recall that the auxiliary fields become rigid parameters after integrating out the three-form sectors.
The additional variations with respect to $\theta(x)$ and the associated three-form yield and additional constraint on the Gauss-Bonnet invariant in terms of the three-form fluxes, $\average{R_\mathrm{GB}} = -\mu^4(\hat{\sigma}'\,\hat{c})/(\sigma'\,c)$, where $c=\integral \varepsilon^{\alpha\beta\gamma\rho}F_{\alpha\beta\gamma\rho}$ and $\hat{c}$ is defined analogously.
Using this additional constraint, it has been shown that vacuum energy contributions can be completely sequestered from gravity, the loops with internal gravitons inclusively~\cite{kalpa6}.

The addition of a coupled Gauss-Bonnet invariant is the minimal modification of the sequestering formalism that also consistently accounts for the pure gravity loops.
Similarly, this term can be invoked in the self-tuning mechanism to prevent the gravitation of vacuum energy beyond the semi-classical limit.
Hence, the most straightforward extension of the original formulation of self-tuning to accomplish that is given by the contribution of $\theta R_\mathrm{GB}$ to the gravitational Lagrangian and the modification of the topological sector of the form
\begin{equation}
\label{eq:34}
    S_{\mathrm{A}} = \frac{1}{4!}\int_{\mathcal{M}}d^4x\, \epsilon^{\mu\nu\rho\sigma}\left[\,\sigma(\varphi)\,F_{\mu\nu\rho\sigma} + \hat{\sigma}(\theta)\,\Hat{F}_{\mu\nu\rho\sigma}\right] \,.
\end{equation}
with variations with respect to $\varphi$, $\theta$, the two coupled three-forms, and the metric. The combination of the two resulting constraints is crucial for preventing vacuum energy contributions from gravitating, with the residual cosmological constant now also being radiatively stable against pure gravity loop corrections.

\subsection{Gauss-Bonnet term in Horndeski gravity}

It is worthwhile to ask whether the sequestering or self-tuning mechanisms in the presence of graviton loops that as described in Sec.~\ref{subsec:invariant} require three or two sets of scalar fields, three-forms, and couplings, respectively, could not be simplified to one set only.
In this respect it is interesting to note that the coupled Gauss-Bonnet term can be cast as a Horndeski scalar-tensor theory~\cite{kobayashi,Jana:2020vov}.
More specifically, let us consider the modification of the action~\eqref{eq:1} by the addition of a non-minimally coupled Gauss-Bonnet term,
\begin{equation}
\label{eq:35}
\begin{split}
    S = \integral \Bigg[\frac{\varphi}{2} R &- V(\varphi)  \\ 
    &+ \theta(\varphi)\,R_\mathrm{GB} + \mathcal{L}_m(\g,\Psi_m)\Bigg] \,,
\end{split}
\end{equation}
where the coupling $\theta(\varphi)$ shall be an arbitrary smooth function of $\varphi$.
It can be shown that the new term in the Lagrangian density, $\theta(\varphi)\,R_{\mathrm{GB}}$, is dynamically equivalent to the following choice of Horndeski terms~\cite{kobayashi},
\begin{align}
    &\mathcal{L}_{\mathrm{GB},\,2} = 8\,\theta^{(4)}\,X^2(3-\ln{X}) \,, \\
    &\mathcal{L}_{\mathrm{GB},\,3} = 4\,\theta^{(3)}\,X(7-\ln{X})\Box\varphi \,,\\
    &\mathcal{L}_{\mathrm{GB},\,4} = 4\,\theta^{(2)}\,X(2-\ln{X})R  \nonumber\\
    &\qquad\qquad\quad + 4\theta^{(2)}(1-\ln{X})\Big[(\Box\varphi)^2+ \varphi_{;\mu\nu}\,\varphi^{;\mu\nu}\Big] \,, \\
    &\mathcal{L}_{\mathrm{GB},\,5} = -4\,\theta^{(1)}\,\ln{X}\,G_{\mu\nu}\,\varphi^{;\mu\nu} \nonumber \\
    &\quad - \frac{2}{3}\theta^{(1)}\frac{1}{X}\Big[(\Box\phi)^3 + 2\varphi_{;\mu}^{~~\nu} \varphi_{;\nu}^{~~\alpha} \varphi_{;\alpha}^{~~\mu} - 3\varphi_{;\mu\nu} \varphi^{;\mu\nu} \Box\varphi\Big] \,, 
\end{align}
where we have defined $\theta^{(n)}\equiv\partial^n\theta/\partial\varphi^n$.
Furthermore, since the sum of Horndeski theories remains a Horndeski theory, Eq.~\eqref{eq:35} can always be cast in the form of Eq.~\eqref{eq:9} for any Horndeski generalization of the scalar-tensor sector in Eq.~\eqref{eq:35}~\cite{kobayashi,armaleo,bruck,Jana:2020vov}.

Importantly, however, some extra care must be taken in the inclusion of the Gauss-Bonnet invariant when the topological sector of Eq.~\eqref{eq:2} is present.
The constraint $\partial_\mu\,\varphi(x) = 0$ it imposes forces the kinetic term $X = -(1/2)\,(\partial_\mu\varphi)^2$ to vanish such that the terms with $\ln{X}$ or $1/X$ may spoil the viability of the theory.
However, cases do exist where the terms $\mathcal{L}_{\mathrm{GB},\,i}$ vanish despite $\partial_\mu\,\varphi(x)\rightarrow 0$.
Whereas terms like $X\,\ln{X}$ and $X^2\,\ln{X}$ vanish trivially in this limit, others involve the product of $\ln{X}$ and $1/X$ with terms that contain second covariant derivatives of the field.
One way to achieve the vanishing of $\mathcal{L}_{\mathrm{GB},\,i}$ is then
to impose an extra condition on the second derivatives such that $\partial_\mu\,\partial_\nu\varphi(x)\rightarrow 0$.
Explicit examples with $\mathcal{L}_{\mathrm{GB},\,i}\rightarrow0$ can then be found, for instance, by adopting the unitary gauge or assuming spherical symmetry.

Hence, we conclude that a Horndeski scalar-tensor theory of a single scalar field amended with the coupled topological sector of Eq.~\eqref{eq:2} can indeed be sufficient to maintain the self-tuning mechanism in the presence of graviton loops.
Importantly, however, this is provided that quantum gravity effects arise in the form of a non-minimally coupled higher-order curvature term, or, in other words, that they have a global dependency on the gravitational coupling. For those certain cases in which the scalar field and the coupling are of adequate form, the quantum effects would then be canceled. The addition of the non-minimally coupled $R_\mathrm{GB}$ term may then be viewed like adding zero, just as for sequestering in Eq.~\eqref{eq:33}. Finally, by integrating out the three-form sector and setting $\varphi\rightarrow\planck{2}$, the effective field equations reduce to those of General Relativity.


\section{Conclusions}\label{sec:conclusions}

The simple additional variation of the General Relativistic Einstein-Hilbert action  with respect to the quadratic Planck mass was recently proposed as a remedy to the Cosmological Constant Problem, and a local scalar-tensor formalism was subsequently developed as counterpart to the originally global framework.
In this paper, we have explored and illuminated further aspects of the self-tuning mechanism arising from the Planck mass variation.

We have first presented a brief review of both the local and global self-tuning formalisms as well as the different aspects of the mechanism that have previously been studied such as the degravitation of the Standard Model vacuum energy, the remedy of the coincidence problem, and the embedding of the local formalism into Horndeski gravity.

Previously, the formalism has only been studied in the Jordan frame, where matter fields follow geodesics of the metric but the metric satisfies a modified Einstein equation.
Here, we have shown that self-tuning operates equivalently in the conformally related Einstein frame, where metrics satisfy the Einstein field equations but do not describe the geodesics of matter fields.
In this process, we have also further clarified and highlighted the similarities and differences between the self-tuning and sequestering mechanisms.
In brief, self-tuning requires one less additional variation of an extra field in the action to operate.
Specifically, in the global approach these are the Planck mass in the Jordan frame or the matter coupling $\lambda$ in Einstein frame.
In the local formalism, self-tuning operates with the variation of the scalar field and the coupled three-form gauge field.
Unlike for sequestering, the cosmological constant is not separately varied, and the attached three-form in the local case is not required.
But despite introducing less auxiliary fields, self-tuning embeds the effective field equations of sequestering for specific forms of the Planck mass or scalar field dependence of the vacuum energy and the classical counterterm.
Importantly, unlike sequestering, self-tuning generally degravitates the vacuum energy not by its cancelation between the geometric and material sides of the Einstein field equations, which only occurs in the limit it effectively recovers the sequestering mechanism, but by the self-tuning of the classical counterterm due to the constraint equation imposed from the extra variations of the action.

We have then extended the self-tuning mechanism to in principle incorporate a dynamical inflationary epoch at early times.
Specifically, by introducing a step function in the three-form sector we allowed the scalar field $\varphi$ to behave dynamically in a regime $\varphi < \planck{2}$ where $\varphi$ remains uncoupled from the three-form.
Once the scalar field reaches the value $\varphi=\planck{2}$, the dynamics freeze by the coupling of $\varphi$ to the three-form sector, which is when the self-tuning mechanism operates.
Importantly, we have shown that due to boundary conditions on the scalar field and the cosmological constant between the dynamical and frozen regimes, the self-tuning and degravitation of the vacuum energy also extends to the dynamical region as there is no fine-tuning that can take place, evading Weinberg's \emph{No-Go} theorem.
The formal extension of the self-tuning action performed to allow for a dynamical early-time epoch furthermore also naturally suggests an alternative spacelike rather than timelike split of the cosmic manifold, opening the door for a multiverse interpretation.
While scalar-tensor models realising inflation and self-tuning in the early and late Cosmos or dynamical and frozen spatial patches in a multiverse can in principle be realised, we have left an analysis of the phenomenological and observational viability of such models to future work.
However, in this context, we have also revisited the anthropic standpoint on the coincidence problem, providing an improved yet still crude estimate for the probability of the  emergence of conscious life in our Universe as a function of cosmic age.
We found to be living at a very typical epoch, where  one should expect the energy densities of the cosmological constant and matter to be of comparable size.
For a dimensionless quantity to compare the emergence of life in the cosmic history of different universes, we introduced the order of magnitude difference of the evolving horizon size of a universe to the size of its protons as the basic building blocks for atoms, molecules, and eventually life.
For our Universe we found a peak at approximately 42 orders of magnitude difference.
An interesting question for future investigation is whether one should expect life to peak at about the same value across the multiverse.

Finally, we have revisited the discussion of whether self-tuning can operate beyond the semi-classical limit, specifically whether the mechanism is resilient to graviton loops.
We have shown that similarly to sequestering, self-tuning operates when the model is supplied with a higher-order curvature invariant coupled to an additional auxiliary field that is moreover coupled to an additional three-form gauge field, which are both varied.
However, while the addition of a higher-order curvature term is inevitable in order to tackle the graviton contributions, we have also shown that for certain cases, the supplementary terms can be recast as a Horndeski scalar-tensor theory of a single scalar field coupled to a three-form.
This, however, crucially depends on the precise form of the quantum gravity corrections, in the sense that they must have a global dependency on the gravitational coupling.

We conclude that the self-tuning mechanism presents a viable solution to both the vacuum and coincidence aspects of the Cosmological Constant Problem that can moreover give rise to a range of new phenomenological features that merit further investigation.


\section*{Acknowledgments}

Parts of this work were conducted in the context of DSB's MSc thesis. LL acknowledges support by a Swiss National Science Foundation Professorship grant (No.~170547). Please contact the authors for access to research materials.

\bibliography{refs}

\end{document}